\documentclass[12pt]{article}

\usepackage{epsfig}
\usepackage{graphicx}
\usepackage{amssymb}
\usepackage{float}

\begin{document}

\title{The wormholes and the Multiverse}

\maketitle

\begin{center}
I.D. Novikov$^{1,2}$, A.A. Shatskiy$^1$, D.I.Novikov$^1$\\
\vspace{1cm}
$^1$Lebedev Physical Institute of the Russian Academy of Sciences,
Astro Space Centre, 84/32 Profsoyuznaya st., Moscow, GSP-7, 117997, Russia.\\
\vspace{0.5cm}
$^2$The Nielse Bohr International Academy, The Nielse Bohr Institute,
Blegdamsvej 17, DK-2100 Copenhagen, Denmark.
\end{center}



\section{Abstract}

In this paper we construct a precise mathematical model of the Multiverse,
consisted of the universes, that are connected with each other by
dynamical wormholes.
We consider spherically symmetric free of matter wormholes. At the same time
separate universes in this model are not necessary spherically symmetric and
can significantly differ from one another. We also analyze a possibility
of the information exchange between different universes.

\section{Introduction}

The general theory of the Multiverse was considered in [1-7]. The mathematical
model of the space-time, that includes multiple elements
have been widely discussed for a quite long period of time. See for
example [8-15].

One of a very first models is a well known Kruskal metric
[8]. There are two identical universes in this model.
They are connected with an
impassable wormhole. This means, that traveling between the universes
is impossible with $v\le c$. Such a model can be modified [11] in such a way, that a
signal with the speed $v\le c$ can pass one way only.
Another type of dynamical wormhole is the
Reissner-Nordstrom metric of eclectically charged black hole [12]. This
particular solution of the Einstein-Maxwell equation gives dynamical wormholes
connecting universes, that are in the past and future with respect to each
other. These are also semi passable wormholes in the
direction from the past to the future. They are also called black and white
holes or "time wormholes". One more type of wormholes in this model is similar
to the Kruskal metric wormhole connecting identical universes.
There are infinite number of universes along the time axis
in this model. It's worth noting that every black and white wormhole
connects two universes in the
past with two universes in the future.
This model can be modified by changing the wormholes structure and topology of
the universes. In such a modified model there are infinite number of
universes along both time and space directions.

The paper is organized in the following way. In section 3 we consider so called
"half closed" universe. Using the results of section 3, we construct the precise
models of the Multiverse in section 4. Each of this models consists of two universes
only. In section 5 we consider so called vacuole model, which is the basis for
the Multiverse model. In section 6 we
make the precise model of
the Multiverse with infinite number of universes. We analyze possible
information exchange between connected universes in section 7. Finally, in
section 8 we summaries our results.

\section{The half closed universe.}

The model of so called half closed universe was first proposed in [13-17].
This
model is the uniform closed Friedman model of the dust universe without a
small spherical part. This model matches the Kruscal metric with the mass equal
to the gravitational mass of the "removed" sphere.
Therefore one can construct the model of almost closed universe
connected by Kruscal wormhole with the infinite asymptotically flat empty
universe.

We use the Tolman's solution [18], that describes the evolution of any spherical
matter distribution with $p=0$ including the empty space $\rho=0$. The interval
can be written as follows:

\begin{eqnarray}
\label{eq:QU}
ds^2=\tilde{a}^2\left[d\tau^2-e^{\lambda(\tau,r)}dr^2- \left(R(\tau,r)\right)^2
(d\theta^2+\sin^2\theta d\varphi^2)\right].
\end{eqnarray}

Here $c=1$, $\tilde{a}$ is a constant with the dimension of length. All other
quantities are dimensionless. The Tolman's solution is:

\[
e^{\lambda(\tau,r)}=\frac{(R')^2}{1+f(r)},\hspace{0.5cm}
\tilde{a}^2 8\pi G\rho =\frac{F'(r)}{R^2R'},
\]
\begin{equation}
\tau +\phi (r)=\frac{\left[ f(r)R^2+F(r)R\right]^{1/2}}{f(r)}+
\frac{F(r)}{\left[-f(r)\right]^{3/2}}
arcsin\left[-R\frac{f(r)}{F(r)}\right]^{1/2},
\end{equation}
where prime means the derivative over $r$ and $f(r)$, $\phi(r)$, $F(r)$ are
arbitrary functions. Suppose, that the region $r<r_0$ filled with matter
represents closed cosmological model. The interval for this model
can be written in the following way:

\begin{equation}
ds^2=dt^2-a^2(t)[dr^2+\sin^2(r)(d\theta^2+\sin^2\theta d\varphi^2)],
\end{equation}
\begin{equation}
\tilde{a}=a_{max}(t),\hspace{0.5cm}a=a_{max}(t)(1-\cos(\eta)),
\hspace{0.5cm}t=a_{max}(t)(\eta-\sin(\eta)),
\end{equation}
\begin{equation}
f(r)=-\sin^2r,\hspace{0.5cm}F(r)=\sin^3r,\hspace{0.5cm}\phi=\pi/2,
\end{equation}
\begin{equation}
a_{max}(t)=\left[8\pi G\rho_{min}/3\right]^{-1/2}.
\end{equation}

If $\pi/2<r_0<\pi$, then we have a model of semi closed universe. Outside the
sphere $r=r_0$, we have:
\begin{equation}
\rho=0\hspace{0.5cm}and\hspace{0.5cm}F'(r)=0\hspace{0.5cm}(r>r_0)
\end{equation}
For this region we choose the origin of time axis in such a way, that
$\dot{R}(0,r)=0$ everywhere when $t=0$, and we choose the scale of the radial coordinate so that when $t=0$ we have $R(0,r)=r_g(r^2+1)$, where $r_g$ is the gravitational
radius for the metric in vacuum. Therefore, the solution in vacuum is
completely defined:
\begin{equation}
f=-(r^2+1)^{-1},\hspace{0.5cm}\tilde{a}F=r_g,\hspace{0.5cm}(r>r_0).
\end{equation}

In order to match the solution (3-6) with the solution in vacuum, one must
set for $r>r_0$:

\begin{equation}
f=-\left[(r+c_1)^2+c_2\right]^{-1}\sin^6r_0,
\end{equation}
\begin{equation}
F=\sin^3r_0,
\end{equation}
\begin{equation}
\phi=\frac{\pi}{2}\left[\frac{(r+c_1)^2+c_2}{\sin^4r_0}\right]^{3/2}.
\end{equation}
Here $c_2=\sin^6r_0$ and $c_1$ should be defined in terms of $r_0$ by applying
the condition of continuity at $r=r_0$. Using (4,8,10) one can find
\begin{equation}
a_{max}\sin^3r_0=r_g.
\end{equation}

Note, that $r_0$ is the size of the wormhole and
$a_{max}$ defines the characteristic scale of the universe.

\section{Precise models of the Mutiverse.}

Using properties of half closed universe it is not difficult to construct
the model of the Multiverse, that consists of two semi closed universes
(Fig 1a). In order to do this,
let us replace asymptotically flat universe by half closed one with the
following metric:

\begin{equation}
ds^2=c^2dt^2-a_1^2(t)\left[dr_1^2+
\sin^2r_1(d\theta^2+\sin^2\theta d\varphi^2)\right],
\end{equation}

Analogously to (6,12) one can write:

\begin{equation}
a_{max1}=\sqrt\frac{3c^2}{8\pi\rho_{min1}},
\end{equation}
\begin{equation}
a_{max1}\sin^3r_{01}=r_{g1}.
\end{equation}

\begin{figure}[H]
\includegraphics[width=0.95\textwidth]{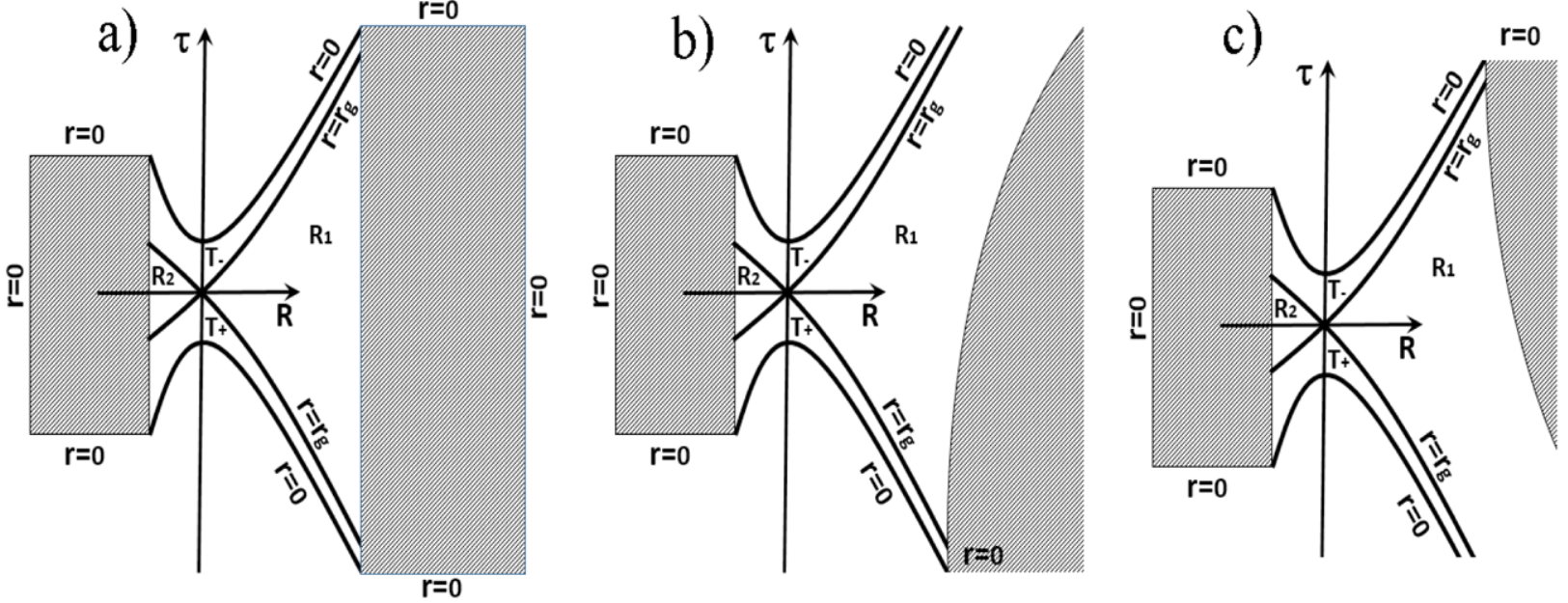}
\caption{The models of the universes connected by the Kruscal wormhole.
a) two half closed universes; b) the half closed universe and
the expanding universe; c) the half closed and the contracting universe.
The shaded area corresponds to the matter of the half closed universe.
$R_1,R_2$ are $R$ regions (see [10]) and $T_+,T_-$ are expanding and
contracting $T$ regions (see [10]). $r_g$ is the gravitational radius.}
\end{figure}

The matching condition is:
\begin{equation}
r_g=r_{g1},
\label{eq_16}\end{equation}
which means, that
\begin{equation}
a_{max}\sin^3r_0=a_{max1}\sin^3r_{01}
\label{eq_17}\end{equation}

Comparing equations (6,12) with (14,15) one can see that two
elements can differ from one another by different $a_{max}$ and
throats size $r_0$.

The case of the particular interest for us is the model
with small $r<<1,\hspace{0.5cm}r_1<<1$, i.e. the
model with $r_g$ much smaller than
the maximum sizes of the universes:
\begin{equation}
a_{max}r^3_0=a_{max1}r^3_{01}.
\end{equation}

Using this approach it is possible to model the Multiverse with
the elements of different types.

Let us consider the combination of half closed universe and the Friedman
expanding universe, that are connected by Kruscal wormhole (Fig 1b).
In both universes $p=0$. The Friedman metric is as follows:

\begin{eqnarray}
ds^2=c^2dt^2-a^2_2\left[dr_2^2+sh^2r_2(d\theta^2+\sin^2\theta d\varphi^2)\right],
\end{eqnarray}

The model is correct for $\Omega<1$, where $\Omega=8\pi\rho_0 G/(3H_0^2)$. Here
$\rho_0$ is the dust density and $H_0$ is the Hubble constant corresponding
to an arbitrary $t_0$. Following [7] one can define the scale factor $a(t)$
by the following way:

\begin{equation}
a_2=a_*(ch(\eta)-1), \hspace{1cm} t=\frac{a_*}{c}(sh(\eta)-\eta),
\end{equation}
Here $a_*$ is the parameter, that actually defines the model in general and
$\eta$ changes from $0$ to $\infty$. Let us isolate a small sphere of the
radius $r_*<<1$ centered at the origin of the coordinate system.
The mass of this sphere is
\begin{equation}
M = \frac{4}{3}\pi\rho R^3 r_*^3 \label{eq_21}\end{equation} Here
$\rho$ is the mass density at some moment $t$, $R$ is a scale
factor at the moment $t$. The whole evolution of the model can be
divided into two periods:\\
(1) When ${t\to t_\infty}$, where $t_\infty$ is the moment of the
beginning of the expansion, and approximately ${\Omega\approx 1}$,
and\\
(2) when ${t\to\infty}$, and ${\Omega << 1}$.\\
For the case (1):
\begin{equation}
\rho = \frac{1}{6\pi G (t-t_\infty)^2},
\label{eq_22}\end{equation}
\begin{equation}
R = R_0\left( \frac{t-t_\infty}{t_0-t_\infty} \right)^{2/3}, \quad
\label{eq_23}\end{equation} $R_0$ is const, and
\begin{equation}
M = \frac{2R_0^3 r_*^3}{9G(t_0-t_\infty)^2} .
\label{eq_24}\end{equation} Here $t_\infty$ and $r_*$ are the
physical parameters of the model.\\
For the case (2)
\begin{equation}
\rho = \rho_0\left(\frac{t_0-t_\infty}{t-t_\infty}\right)^{3},
\label{eq_25}\end{equation}
\begin{equation}
R = R_0\left( \frac{t-t_\infty}{t_0-t_\infty} \right),
\label{eq_26}\end{equation} and
\begin{equation}
M = \frac{4}{3}\pi\rho R_0^3 r_*^3 .
\label{eq_27}\end{equation}
Here $\rho_0$ and $r_*$ are the physical parameters.

One matching condition analogously (\ref{eq_16}, \ref{eq_17}) is
\begin{equation}
a_{max1} r_1^3 = \frac{2GM}{c^2}
\label{eq_28}\end{equation}
Therefore, analogously to the previous case, the sizes of the wormholes
could be completely different.

On more type of the Multiverse is the model where half closed universe
connected with the open contracting one (Fig 1c). To construct such a model
it is enough to change the sign of Friedman parameter $\eta$ in (Eq. 20).

Finally it is possible to consider the model, where the open
expanding universe is connected with analogous one or the model, where
the open expanding Friedman universe connected with the open contracting
one.

Another approach to the Multiverse modeling see in [6].

In section 7 we analyze possible information exchange between two
universes for various cases.

\section{The "vacuole" model of the universes.}

In previous section we restricted our analysis by the models with
two universes only (except for the Reisner-Nordstrem model).
Let us generalize the problem and construct the
Multiverse with infinite number of universes. To do this we analyze
so called "vacuole" model [19-21].

Let us start with the uniform expanding Friedman model with zero pressure.
This could be ether open or closed universe. We define the sphere of the
radius $r_0$ and compress the matter inside this sphere to the compact
object in it's center. According to the equations of the General
Relativity, the gravitational field outside such a "vacuole" will not
change. Therefore the existence of such a vacuole will not affect the
matter expansion outside it.
It can be shown by using the Tolman's solution [18] with $p=0$.

The solution inside the vacuole corresponds to the Schwarzschild solution
written in the expanding coordinate system and the solution outside the
vacuole corresponds to the Friedman solution.

One can obviously make an arbitrary number of vacuoles in Friedman universe.
The only necessary condition is that they shouldn't overlap.

\section{The precise model of the Multiverse with the infinite number
of universes.}

In order to construct the model of the Multiverse using the "multivacuole"
Friedman model, we should simply replace compact objects inside each vacuole by
Kruscal wormholes with corresponding masses. These wormholes should connect
the universe with other universes of arbitrary types. This construction can
be repeated for each universe, connected with the first one. Other universes
can be connected with one another in the same way. Therefore, one
can continue this process and it is possible to connect arbitrary
(up to infinite) number of universes with
each other by arbitrary number of wormholes.

\section{Possible signal exchange between the universes.}

The detailed analysis of the radiation transfer in Kruscal metric has been
done in [22,23]. In case of empty universes, connected by Kruscal wormhole,
it is impossible for
the signal with speed $v\le c$ to propagate from one universe to another.

The equation for the radial signal propagating at the speed of
light can be written as follows:

\begin{equation}
\left(\frac{dr}{dt}\right)^2=-\frac{g_{00}}{g_{11}}
\end{equation}

In Fig (2a) one can see, that the signal can pass from one universe
to another (from A to B) only at the very beginning of the evolution
of the universe A. In other words, for the signal to propagate from A to B it
should exit from A when the edge 5 is in the region $T_+$, i.e.
between 4 and 7, or when the edge of B is in the region $T_-$, i.e.
between 4 and 10. The situation is obviously the same for the signal,
passing from B to A.

Fig (2b) shows the case for the half closed (A) and the expanding (B)
universes. The signal can propagate from A to B only if it exits from A
at the moment between 4 and 7. Fig (2c) shows the analogous case, but
for the combination of half closed and contracted universes. Finally
Fig (2d) shows the combination of two expanding universes. It is
possible also to get the result for other combinations of universes.

\begin{figure}[H]
\includegraphics[width=0.95\textwidth]{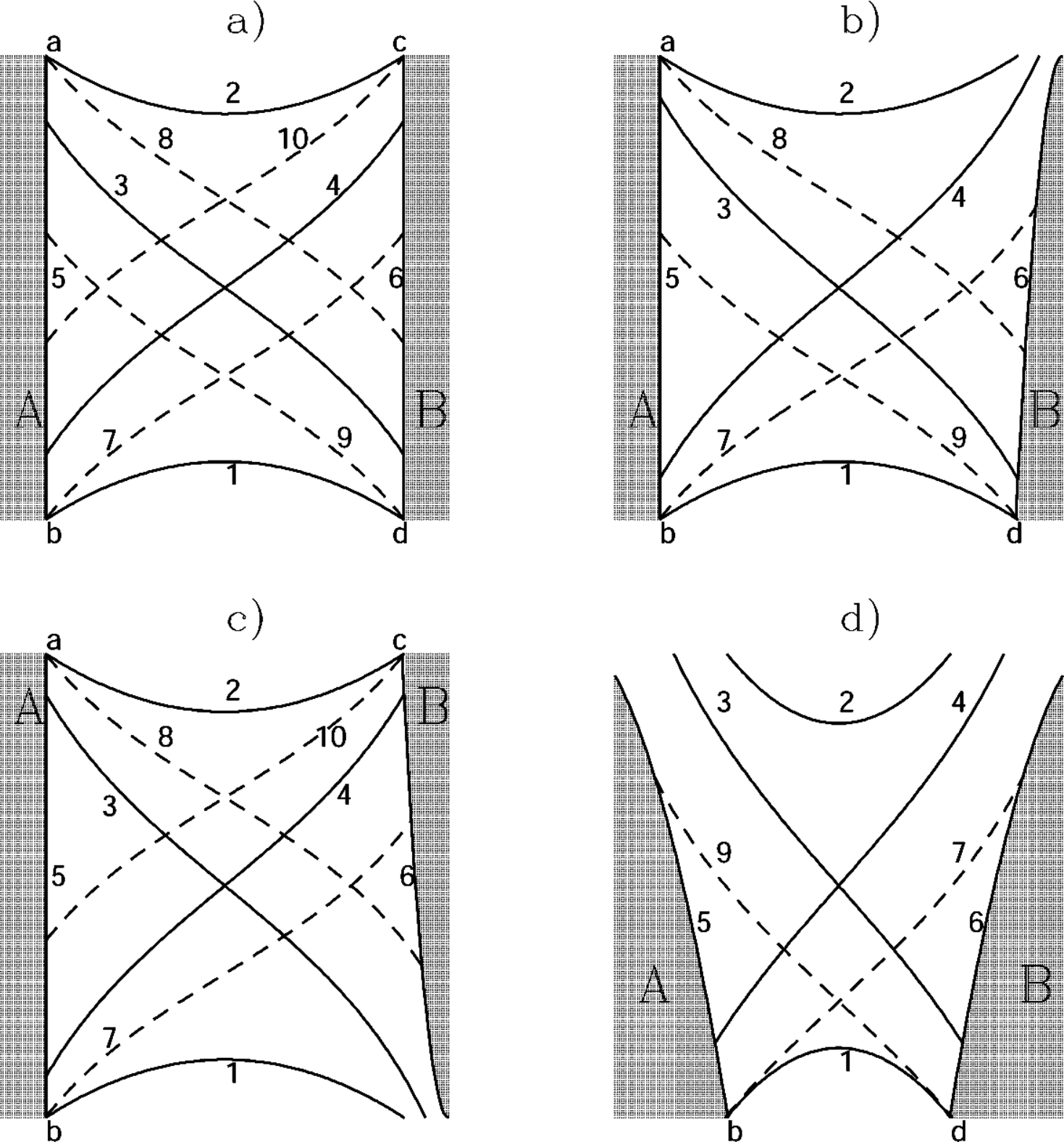}
\caption{The information exchange between two universes. a) Two
half closed universes A and B; b) The half closed (A) and the
expanding (B) universes; c) The half closed (A) and contracting
(B) universes. Lines 1 and 2 correspond to singularities, 3 and 4
correspond to the gravitational radius $r_g$, 5 and 6 are the
edges of A and B respectively. The line 7 is the zero geodesic
line, coming from the point b. 8 is the zero geodesic coming to
the point a. 9 is the same as 7, but coming from d, and 10 is the
same as 8, but coming to c.}
\end{figure}

Possible application of our theoretical results to an observations suggests,
that the most interesting case is the model of the expanding universe
Fig (2b,d).
The signal in such a model can come to B if it propagates between lines
4 and 7. The very first signal, that observer in the universe B can see, is
the signal propagating along the line 7 from point "b", that corresponds
to the origin of the universe A. It comes from singularity and should
correspond to the sharp outbreak [24].

The parabolic expansion of the sphere from the white hole horizon $r_g$
and how does it look like for the stationary observer in Schwarzschild
reference frame was discussed in [22,25]. Let us apply this result for the
expanding sphere from the universe A and the observer in B. The sphere
starts expanding from the edge of A between lines 7 and 3 and comes
to $r_g$ (line 4, see Fig (2)). If the observer at the edge of B
observes the radiation from this sphere late enough, then the speed of
the observer in Schwarzschild reference frame is small and we can
consider this observer far from the white hole as a stationary one.
The infinitely remote observer should see the first signal from the edge
of A, that originated at the beginning of the sphere expansion. According
to [22] in $\Delta t\approx 0.28r_g/c$ after the first signal, the observer
should see the radiation in the center
of the visible disc, that escaped the expanding sphere at
the moment, when this sphere crossed the Schwarzschild sphere $r_g$.
The observed frequency of this radiation is twice as much as the
initial one. The angular size of the visible disc is
$\varphi\approx 0.43r_g/r$, where $r$ is the coordinate of the observer in the
Schwarzschild reference frame.

\section{Conclusions.}

In this paper we considered possible connections between
different universes in the Multiverse with $p=0$.
Using the so called "vacuole" model we constructed the model of the Multiverse
that consists of many universes (up to infinite number of them). We
demonstrated, that these universes can have different properties.
We analyzed the possibility of information exchange between the
universes and showed at what epochs such an exchange is possible.  We remember that in the case of the Kruskall metric the wormhole is impossible. In the case  of the Kruskal tiye of the wormhole connected the universes with the matter these wormhole are possible.

It's worth noticing, that it is possible to
generalize this model for $p\ne 0$. In this case the vacuole model is
correct for the period of time when the rarefaction wave propagates from
the edge of the vacuole to its centre, see [19,24].

The investigation of possible instabilities should be done separately,
see [22]. Another aspects of the problem see in [26].

\begin{center}
{\it Acknowledgments}

This work was supported by the RFFI Foundation with the project codes 12-02-00276-a,
13-02-00757-a and by the Scientific School 14.120.14.4235-NSH.
\end{center}

\newpage

\begin{center}
{\it References}
\end{center}
1. B. Carr, "Universe or Multiverse?" (Cambridge Univ. Press, 2009).\\
2. A.D. Linde, "Particle Physics and Inflationary Cosmology" (Harwood, Chur,
Switzerland, 1990).\\
3. K.A. Bronnikov, Acta Phys. Polon. B, 4, 251 (1973).\\
4. I.D. Novikov, A.G. Doroshkevich, J. Hansen and A.Shatskiy, Int. J. Mod. Phys.
D 18, 1665 (2009).\\
5. M. Visser, "Lorentzian Wormholes: from Einstein to Hawking",
(AIP, Woodbury, 1995).\\
6.Hideki Maeda, Tomohiro Harada and B.J. Carr (2009), ArXiv: 0901.1153.\\
7. C.W. Misner, K.S. Thorn and J.A. Wheeler, Gravitation,
(W.H. Freeman \& Co., San Francisco, 1973).\\
8. M.D. Kruskal, Phys. Rev. 119, 1743 (1960).\\
9. I.D. Novikov, Astr. Zh, 40, 772 (1963).\\
10. I.D. Novikov, Publications of the Shternberg Astronomical Institute, Moscow State University,
132, 3, 43 (1964).\\
11. I.D. Novikov, A.G. Doroshkevich, A.A. Shatskiy, D.I. Novikov, Astr. Zh, 86,
1155 (2009).\\
12. S.W. Hawking, G.F. Ellis, "The large scale structure of space time",
(Cambridge Univ. Press, Cambridge 1973).\\
13. O.Klein, Werner Heisenberg und die Physic unserer Zeit, s. 58,
(Braunschweig: Keweg) (1961).\\
14. Ya.B. Zeldovich, Zh. Eksp.Teoret. Fis., 43, 1032 (1962).\\
15. I.D. Novikov, Vestnik Moscow State University ser. 3, 6, 61 (1962).\\
16. R.K. Harrison, K.S. Thorne, M. Wakano, J.A. Wheeler, Gravitation Theory and
Gravitational Collapse (Chicago Univ. Press, 1965).\\
17. Ya.B. Zeldovich and I.D. Novikov, Relativistic Astrophysics (Moscow, NAUKA Press, 1967).\\
18.  R.C. Tolman, Phys. Rev. 35, 875 (1930).\\
19. A. Einstein and E.G. Straus, Rev. Mod. Phys. 17, 120 (1945).\\
20. I. Novikov, S. Fillipi, R. Ruffini, Lettere al Nuovo Cimento 39, 185 (1984).\\
21. I.D. Novikov, Astr. Zh. 41, 1075 (1964).\\
22. V.P. Frolov, I.D. Novikov, Black Hole Physics (Kluwer Academic Publishers, 1998).\\
23. I.D. Novikov, L.M. Ozernoj, Doklady Acad. Nauk SSSR, 150, 1019 (1963).\\
24. A. Retter and S. Heller, ArXiv. 1105.2776, (2011).\\
25. Ya.B. Zeldovich, I.D. Novikov, Relativistic Astrophysics, vol.2
(Chicago Univ. Press, 1983).\\
26. I.D. Novikov, A.A. Shatskiy, D.I. Novikov, Astr. Zh., (2015), in press.

\end{document}